# Magnetic properties of ferromagnetic $Gd_5Si_4$ - $Fe_3O_4$ bilayer thin film heterostructure

Shivakumar Hunagund, V Santiago, C Castano

*Department of Mechanical and Nuclear Engineering, Virginia Commonwealth University, Richmond VA 23284 USA*

## Abstract

Magnetic properties of thin films differs from that of the corresponding bulk materials and show novel physics as a result of their reduced size and dimensionality. Profound changes in magnetic properties can also result with exchange bias between the interface of the thin films of different materials. Here we present $Fe_3O_4$, $Gd_5Si_4$ and $Fe_3O_4$-$Gd_5Si_4$ bilayer thin films of varying thicknesses were prepared by RF Magnetron sputtering of $Fe_3O_4$ and $Gd_5Si_4$ targets on silicon substrates in high vacuum. Their magnetization was measured at room temperature in fields of 3T and between 50 K to 400 K temperatures. $Gd_5Si_4$ films resulted in oxidation due to the exposure to the air. Deposited film thickness of all the specimens are larger than 50nm and show ferromagnetic behavior.

## Introduction

Thin films are used in wide variety of high technology and industrial applications like data storage, batteries, sensors and microelectronics. There has been significant developments in magnetic thin films materials and fabrication in recent years. Thin films can be deposited with different processes such as spraying, spin-coating, dip-coating, chemical vapor deposition (CVD), evaporation, and sputtering.

Intrinsic magnetic properties such as magnetic saturation (Ms), Curie temperature (Tc) etc can be significantly different in thin films and in the bulk due to finite size effect [1]. Furthermore, it has been observed an exchange interaction results at the interface between a ferromagnetic material and an anti-ferromagnetic / ferrimagnetic material [2][3]. The exchange bias appears when the curie temperature of the ferromagnet is above the Neel temperature of the anti-ferromagnet / ferrimagnet [2]. This phenomena is exploited in data storage technology. In this paper we present deposition of $Fe_3O_4$ and $Gd_5Si_4$ films using RF Magnetron sputtering. Although $Fe_3O_4$ deposition with RF/DC sputtering has been widely reported, $Gd_5Si_4$ thin film deposition with this method has never been reported thus far.

$Fe_3O_4$ has an inverse spinel crystal structure with two cation sites- $Fe^{2+}$ and $Fe^{3+}$ ions that are encircled by oxygen ions to form tetrahedra and octahedra structures. $Fe_3O_4$ is ferrimagnetic whose magnetic property is influenced by the $Fe^{2+}$ and $Fe^{3+}$ ions as the spin of these ions at these two cation sites are anti-parallel coupled by super-exchange effect. $Fe_3O_4$ is reported to exhibit high Curie temperature (~860 K) [4]. $Gd_5Si_4$ is ferromagnetic material with a reported transition temperature for PLD deposited thin film at 342 K [5]. It has an orthorhombic structure [6].

In this paper we report deposited thin films of varying thicknesses of $Fe_3O_4$, $Gd_5Si_4$ and $Fe_3O_4$-$Gd_5Si_4$ bilayer present their detailed characterization and analysis.

## Methods

The samples were deposited on silicon substrate using RF magnetron sputtering using a pure $Fe_3O_4$ and $Gd_5Si_4$ ceramic targets. The apparatus is lab assembled unit equipped with RF generator, and a pumping system composed of a mechanical pump coupled with a turbo molecular pump. The base pressure of the growth chamber was on the order of *4 x 10$^{-6}$* Torr. The apparatus is lab assembled unit equipped with RF generator, and a pumping system composed of a mechanical pump coupled with a turbo molecular pump. The distance between the target and the substrates was 55 mm and the RF power supply was set at 80 W for $Fe_3O_4$ deposition and 40 W for $Gd_5Si_4$ deposition. RF bias power densities of 0 W/cm2 were applied to the substrate. The gas used in this study was argon and the working pressure was kept at a value of 12 mTorr and 37 mTorr respectively. The flow was kept constant at a rate of 20 sccm for $Fe_3O_4$ and 25 sccm for $Gd_5Si_4$. Initial deposit was made with $Fe_3O_4$ followed by a top layer of $Gd_5Si_4$ to complete the bilayer structure. The surface roughness of this both layers was measured by atomic force microscopy. Furthermore, five samples of varying thickness of monolayers were deposited separately: 3 $Fe_3O_4$ films of deposition times of 15 minutes (*S1*), 30 minutes (*S2*) and 60 minutes (*S3*) and 2 $Fe_3O_4$ - $Gd_5Si_4$ bilayer films of deposition times of 30 minutes for $Fe_3O_4$ and 1 hour of $Gd_5Si_4$ (*S4*) and 1 hour each of $Fe_3O_4$ followed by $Gd_5Si_4$ (*S5*). Film thicknesses were measured using a Zeiss Auriga Crossbeam FIB - SEM dual system. Part of the film was removed from top down to the substrate and the film thickness was thus measured with the sample slanted. Phase composition of the films were studied with PHI VersaProbe III Scanning XPS system. Morphology and microstructure of the as-deposited samples were examined by Dimension FastScan atomic force microscopy (AFM) and Hitachi SU-70 S/TEM. Finally, Magnetic properties were characterized with Quantum design Physical Property Measurement System (PPMS) system.

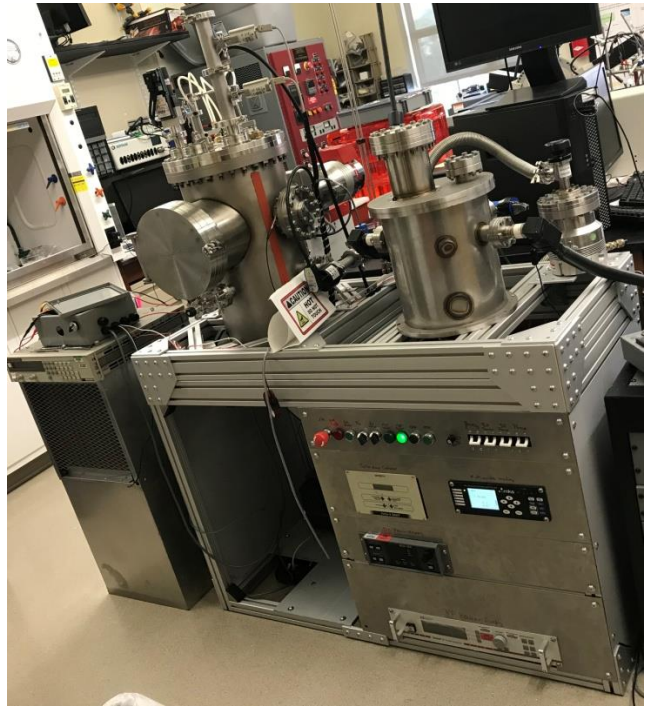

*Fig.1* RF / DC Magnetron system used for sputter deposition of the films.

## Results and discussion

To observe the thicknesses, the samples were first coated with Pt in order to protect the specimens surface during milling. Then the specimens were milled to upto 5 µm depth using a Zeiss Auriga focused ion beam (FIB). The FIB milling voltage is set to 2 kV to minimize damage from implanted Ga. Specimens observed under SEM shows (*Fig. 1*) films thicknesses of 58 nm for *S1*, 96 nm for *S2*, 228.91 nm for *S3*, 83 nm of $Fe_3O_4$ and 90 nm of $Gd_5Si_4$ for *S4* and finally 110 nm of $Fe_3O_4$ and 175 nm of $Gd_5Si_4$ for S5. The thickness of the films are linearly proportional to the deposition times provided other parameters in the RF Magnetron sputtering remains the same.

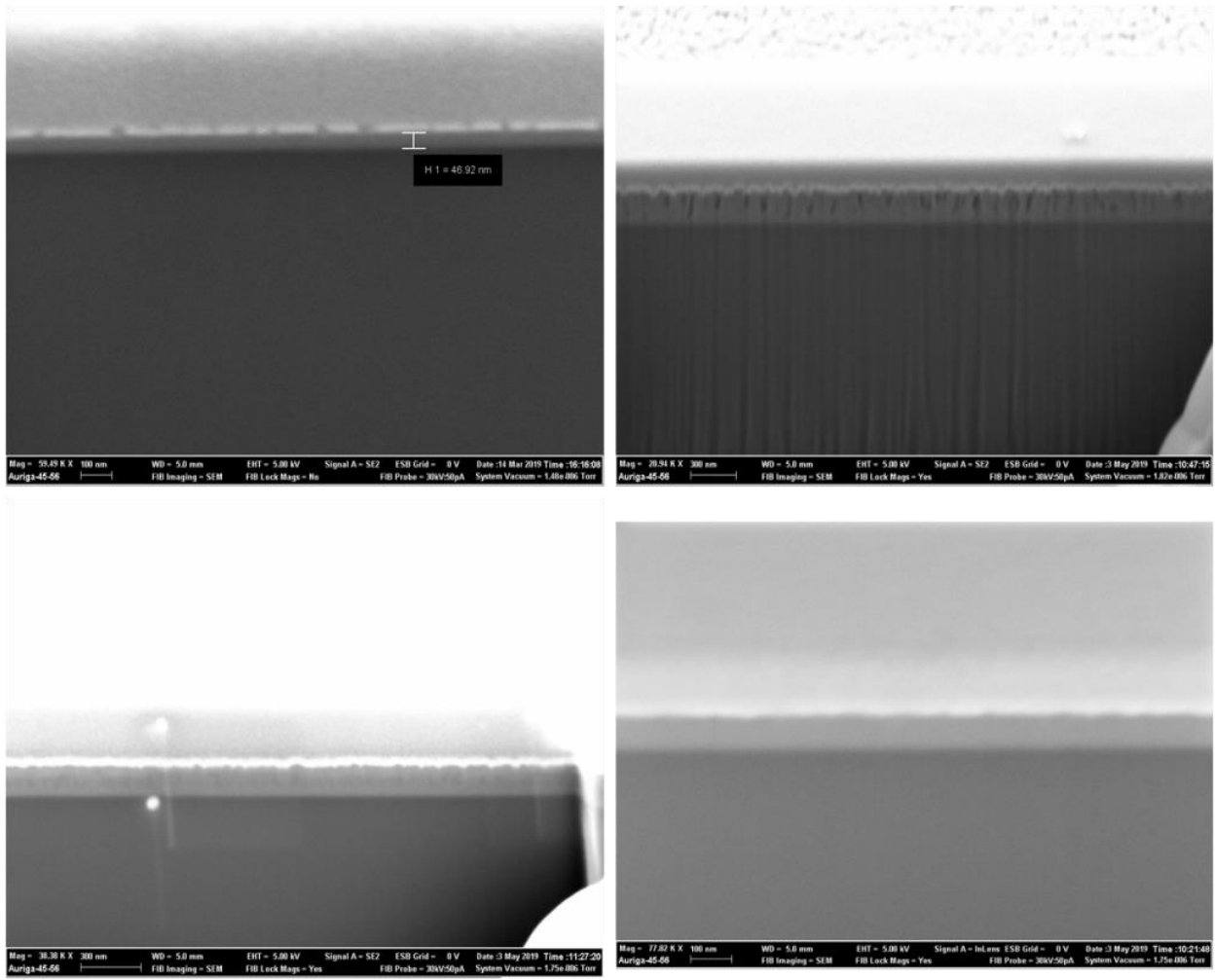

Fig. 2 SEM images of the specimen S1 - S4 (viewed at 54° angle). The bright top layer is the protective Pt layer and the darker bottom layer is the silicon substrate.

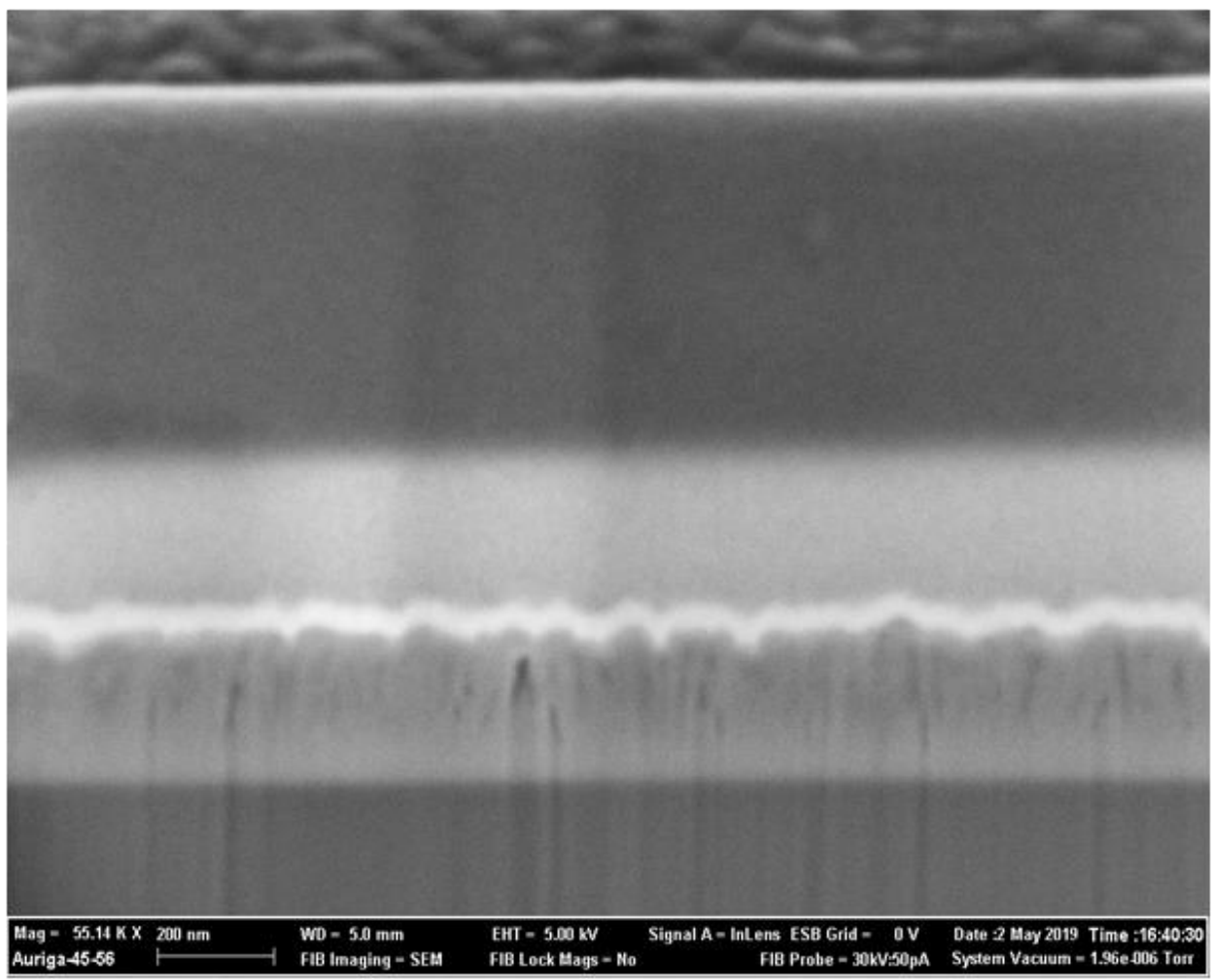

Fig. 3 SEM image of the specimen S5. Note contrast visible at the interface between the Pt layer on top and the silicon substrate. $Gd_5Si_4$ is deposited on $Fe_3O_4$.

The roughness of the both the $Fe_3O_4$ and $Gd_5Si_4$ film surface was found to be 0.738 nm, and 3.02 nm, respectively, as measured by atomic force microscopy (AFM), shown in *Fig. 4 and Fig. 5*. Silicon wafer used as the substrate in this experiment and its surface is regarded to be "flat". Interfacial topographies and surface morphology observed in SEM and AFM reveals non-homogeneous thickness in films and also significant void fraction in Gd5Si4 film. During growth, stress relaxation in between film interfaces strongly alters growth characteristics of the following film deposition [10]. The respective surface roughness esp. at the interfacial layer will influence the global magnetic interaction as the physical contact area between two rough surface will depend on the roughness. Therefore, the amount of disorder at the surface/interface can also influences magnetic properties of thin magnetic films, such as coercivity, magnetic domain structure, and magnetization reversal. Also, surface/interface roughness has been shown to have a significant influence on the demagnetizing field [11][12].

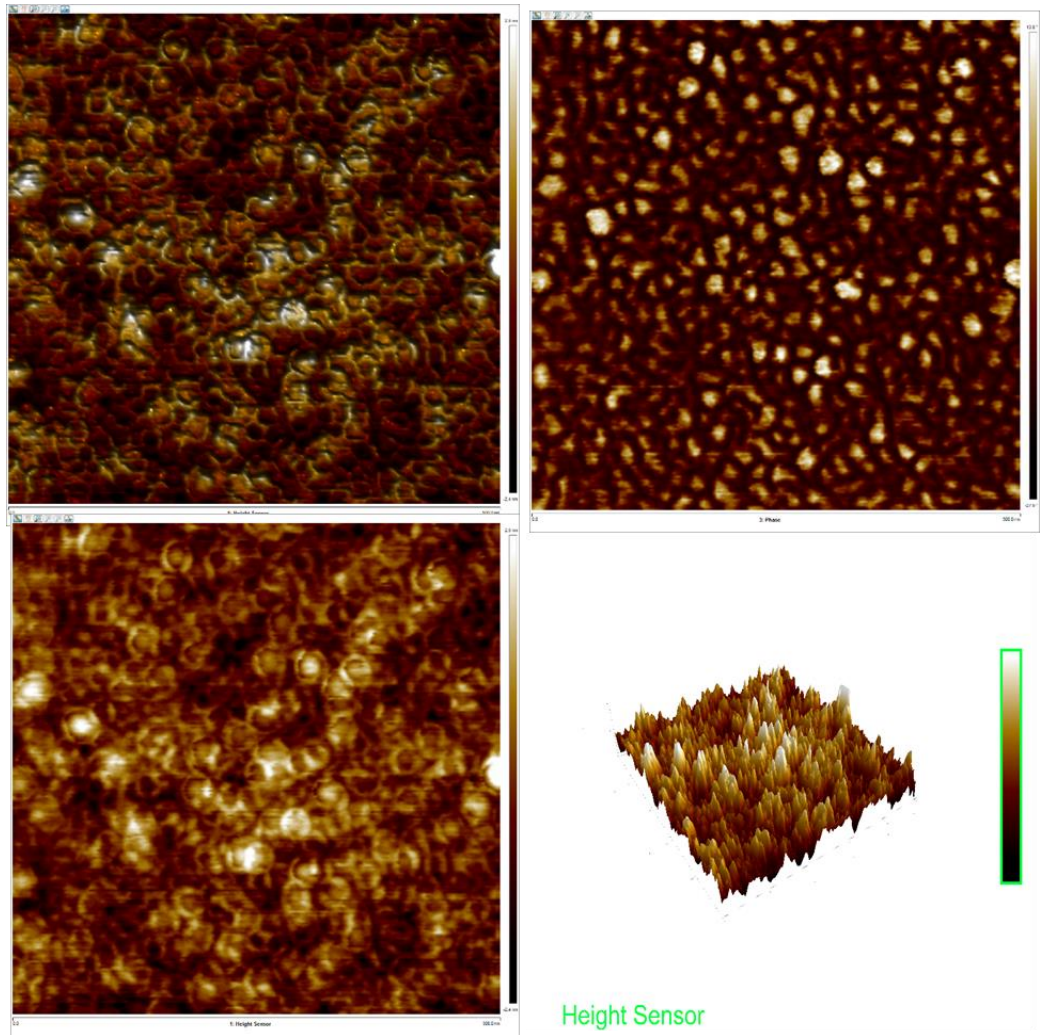

Fig. 4 AFM images of the Fe$_3$O$_4$ thin film shows surface roughness of **0.738 nm**. The surface roughness was analyzed with Nanoscope analysis software.

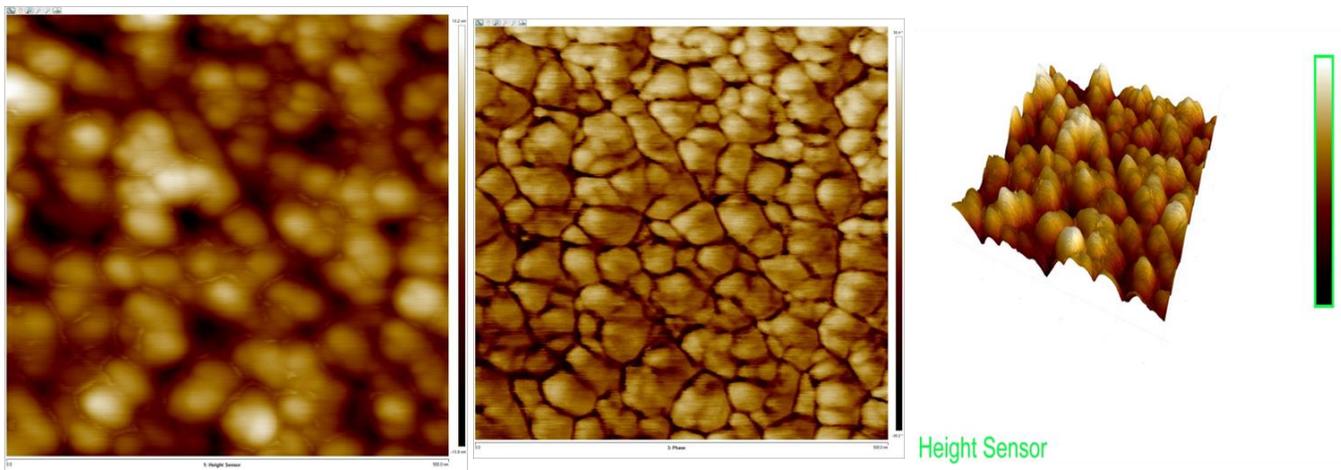

Fig. 4 AFM images (Height, Phase and 3D) of the $Fe_3O_4$ - $Gd_5Si_4$ thin film. Analysis shows surface roughness of 3.**02 nm**.

XPS surface characterization analysis shows elemental composition and phases. The XPS survey scan of the $Fe_3O_4$ thin film deposited on Si (1 0 0) substrate is shown in Fig. 5 , we have observed that only Fe and O are present with very small contribution of C which was expected because film surface was exposed to air before XPS measurements. The positions of various photoemission peaks are marked in the spectrum for elements present in the film. Further detailed scan have performed for Fe 2p core level spectra to determine charge/electronic state of elements present in the film.

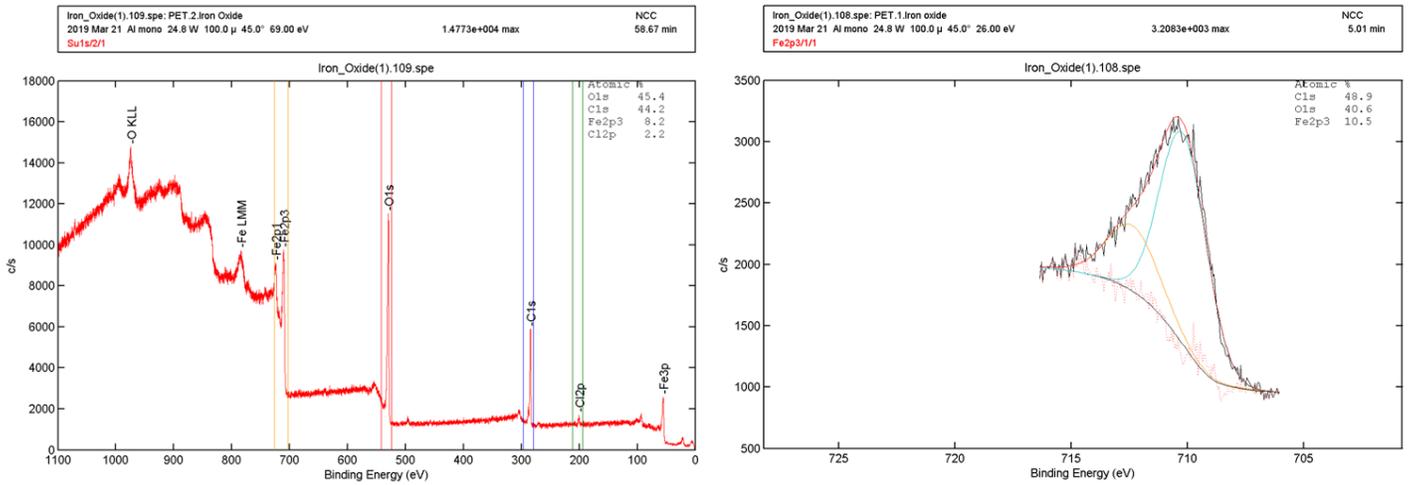

Fig. 5 XPS Survey spectrum of $Fe_3O_4$ film indicating presence of carbon. Analysis reveals presence of phases other than $Fe_3O_4$.  (*right*) $Fe_3O_4$ (XPS SPECTRUM) - Region: Fe2p3 spectrum curve fitted to identify peaks.

Fig. 5 depicts the high-resolution scan of the Fe 2p core level. The deconvoluted spectrum shows the presence of two peaks at 710.18 eV attributed to $Fe_3O_4$ phase and 712.35 eV.

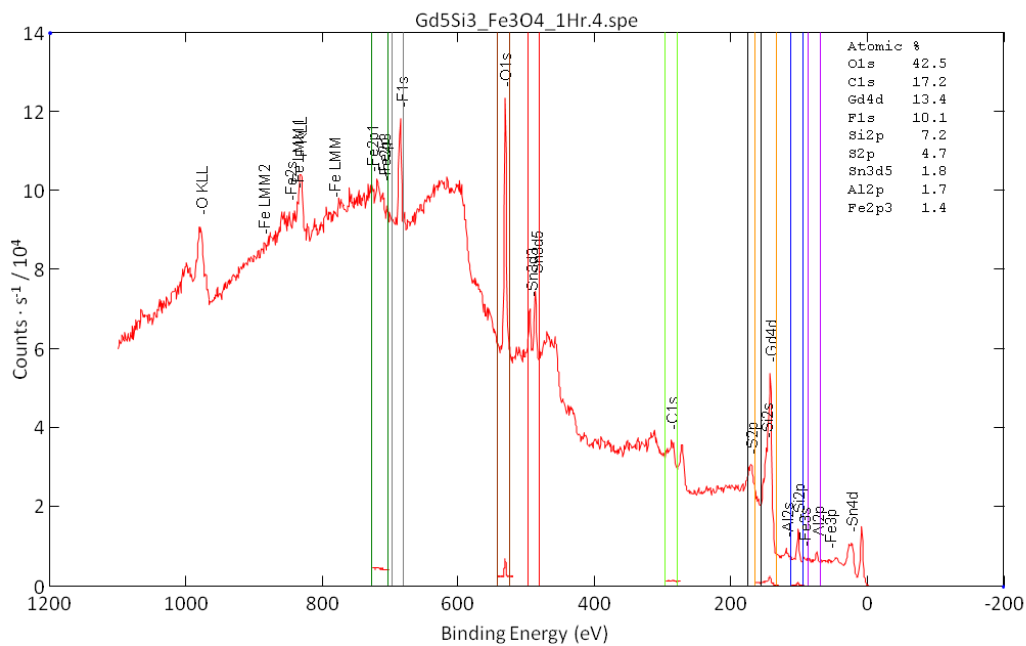

Fig. 6 XPS quantitative analysis of the Survey spectrum of $Fe_3O_4$ - $Gd_5Si_4$ film elemental contents of the thin film sample. Analysis reveals presence of Gd, Fe, O, Si, and F. Prior to XPS measurements, the film was plasma treated to remove any contamination of the surfaces.

The survey scan of the $Gd_5Si_4$ deposited on $Fe_3O_4$ is shown in Fig. 6 Gd 4d core-level XPS spectrum of thin film is shown in Fig. 7. The features of Gd are fitted with Gaussian–Lorentzian functions which reveals binding energies of 142.50 eV, 147.98 eV and 152.69 eV. O 1s core level spectra has binding energy position at around 531.42 eVand for the Si 2p binding energy at 101.51 eV. All XPS core level spectra were fitted using PHI MultiPak data reduction and interpretation software package.

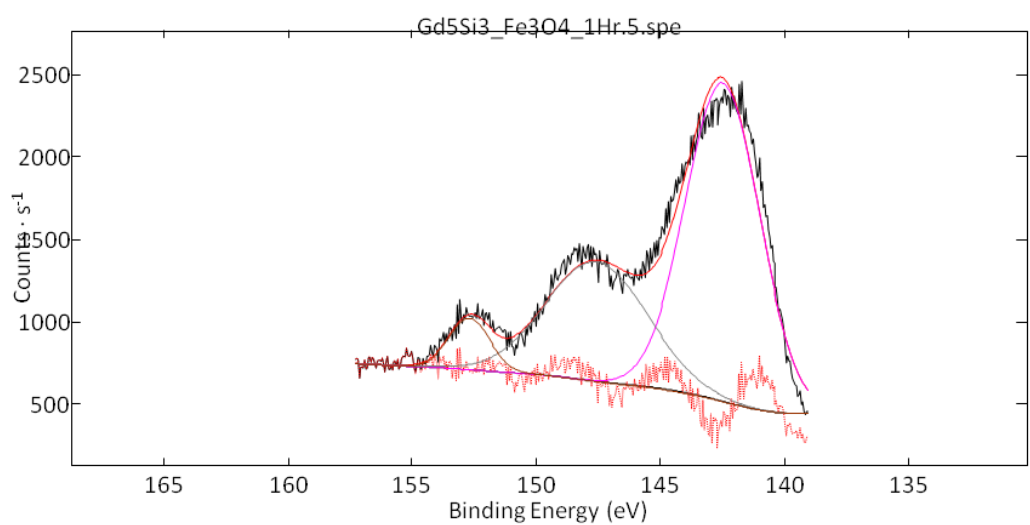

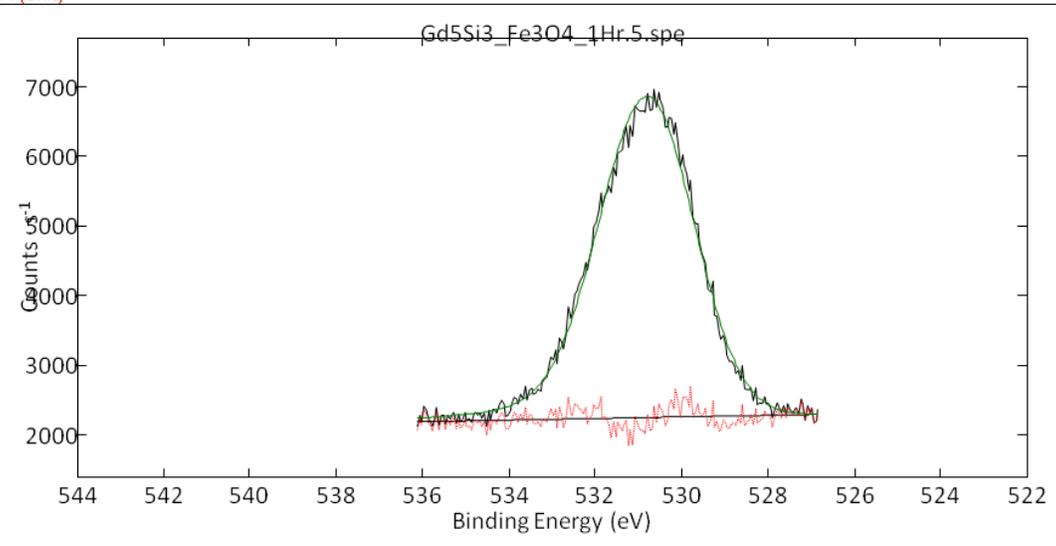

Fig. 7 XPS SPECTRUM - Region: Gd4d/3 spectrum curve fitted to identify peaks. (*bottom*) Region: O1s/3 spectrum curve fitted to identify peaks.

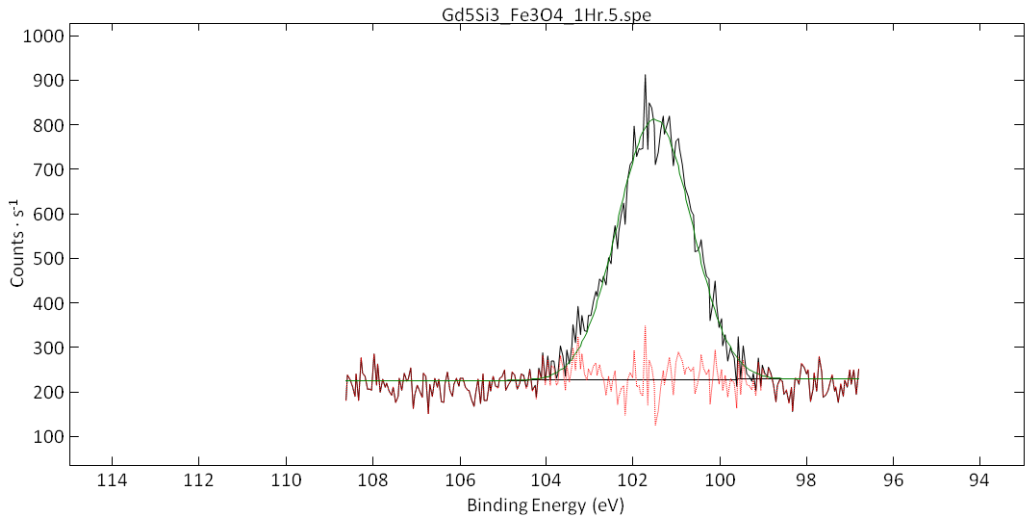

Fig. 8 XPS SPECTRUM - Region: Si2p/3 spectrum curve fitted to identify peaks.

The hysteresis (M-H) curves for the $Fe_3O_4$ thin films and $Fe_3O_4/Gd_5Si_4$ bilayer films are shown in Fig. 8, 9, 10, 11 and 12. The magnetic hysteresis plots of film at the room temperature and 50 K to 400 K were measured using QD PPMS. It can be seen that the $Fe_3O_4$ film deposited on glass substrate shows higher saturation magnetization (Ms = 8 emu), while the films deposited on Si(100) show lower saturation magnetizations (Ms = 2.6 emu for S1 and Ms = 2.9 emu for S2).

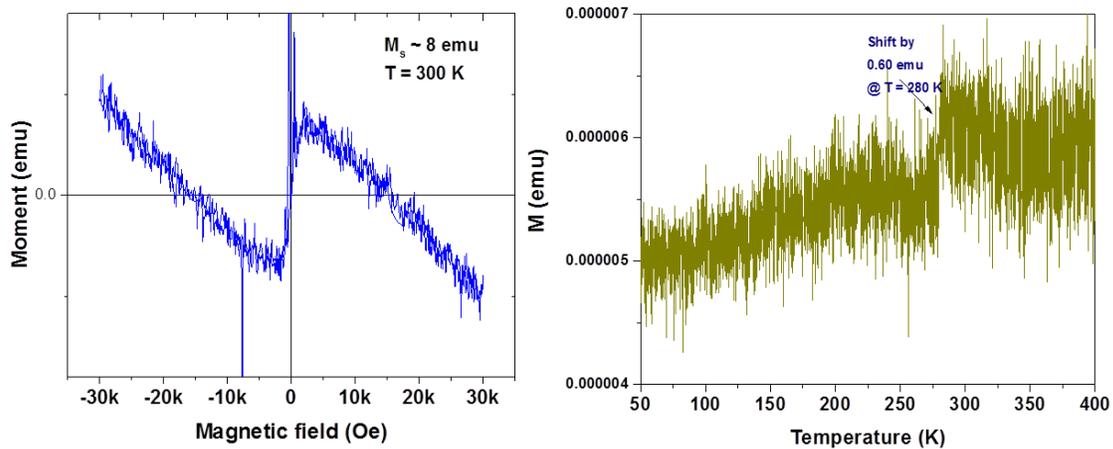

Fig. 9 $Fe_3O_4$ (S1) thin film exhibit ferromagnetic behavior with magnetic saturation at around 8 emu. The diamagnetic features in observed is due to glass/silicon substrate. (*right*) M-T Curve shows shift in moment by 0.60 emu at the temperature of 280K indicating possible transition.

This is because S1 is deposited on glass substrate while the S2 and S3 are deposited on Si(100), hence nature of the substrate surface plays a vital role in the synthesis of thin films and the inherent characteristics of substrate are very important [4][16]. Surface properties of glass and silicon have distinct influence on nucleation and growth processes of thin films [14]. There is large lattice mismatch of substrates (Si, a = 5.404 Å) and films ($Fe_3O_4$, a = 8.397 Å). The curie temperature (Tc) was obtained from differentiating magnetic moments with respect to the temperature. Tc for $Fe_3O_4$ specimens were observed at 281 K for S1, 378 K for S2 and 300 K for S3.

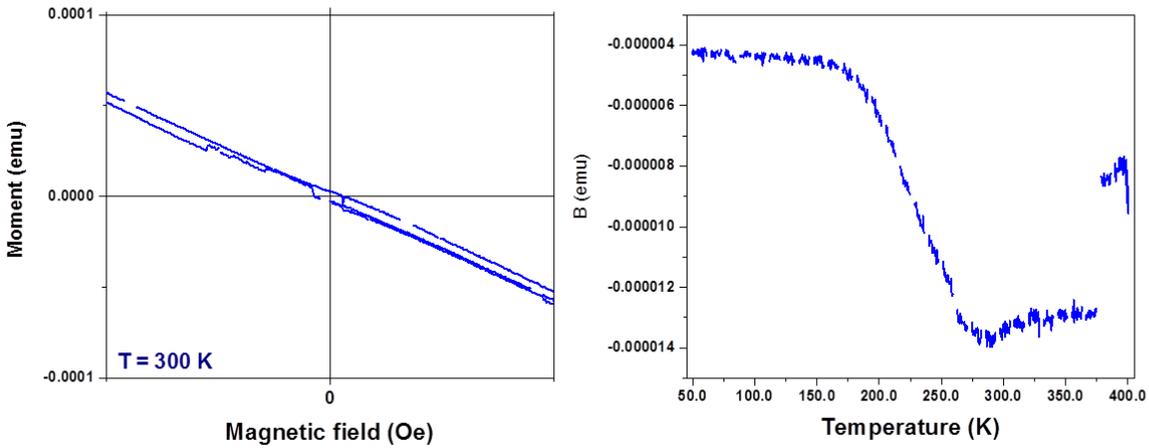

Fig. 10 M-H curve of the $Fe_3O_4$ (S2) thin film showing magnetic saturation (Ms) at about 2.6 emu and Tc at 378 K.

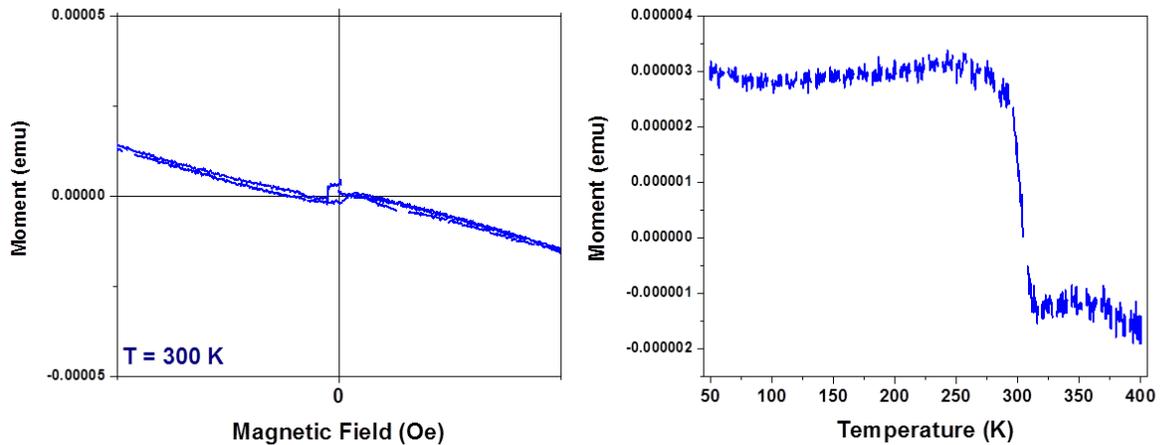

Fig. 11 M-H curve of the $Fe_3O_4$ (S3) thin film showing magnetic saturation (Ms) at about 2.9 emu and Tc at about 300 K.

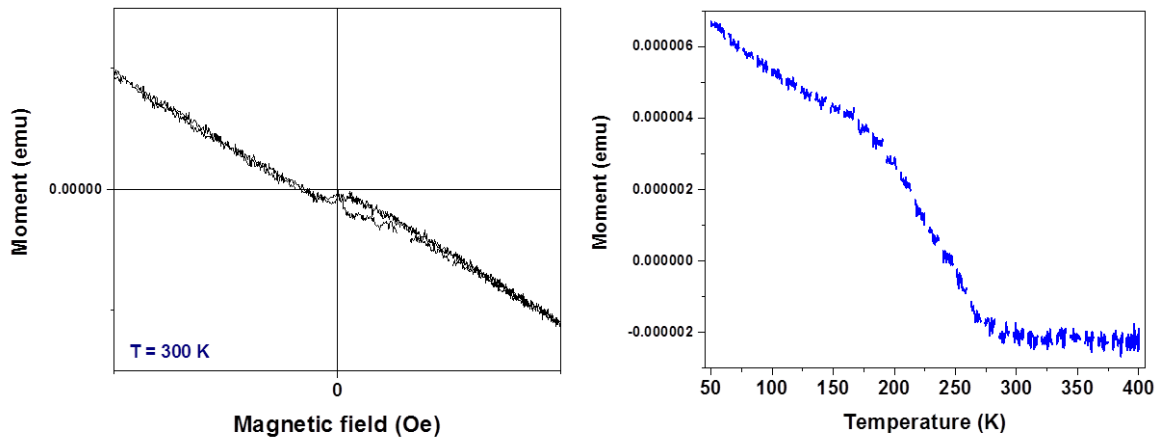

Fig. 12 M-H curve of the Gd$_5$Si$_4$ - Fe$_3$O$_4$ (S4) bilayer thin film heterostructure retains ferromagnetic behavior. The magnetic saturation (Ms) is about 1.58 emu. Transition temperatures are observed at around 148 K, 240 K and 300 K. Note Gd$_5$Si$_4$ has Tc = 318 K and Fe$_3$O$_4$ has Tc = 858 K.

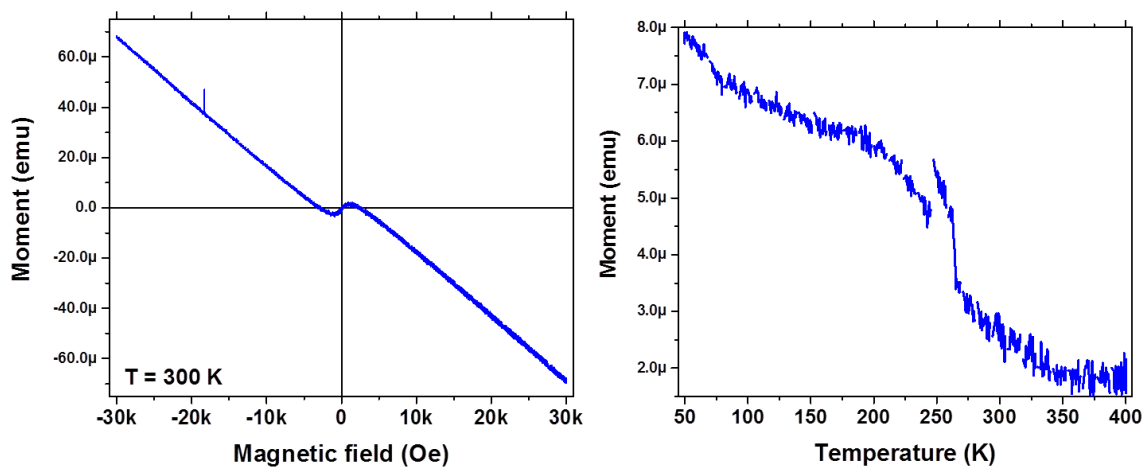

Fig. 13 M-H curve of the Gd$_5$Si$_4$ - Fe$_3$O$_4$ (S5) bilayer thin film heterostructure retains ferromagnetic behavior. The magnetic saturation (Ms) is about 3 emu. Transition temperatures are observed at around 243 K and 265 K. Note Gd$_5$Si$_4$ has Tc = 318 K and Fe$_3$O$_4$ has Tc = 858 K.

Specimen S4 exhibited Tc at around 148 K, 240 K and 300 K while for S5 Tc is observed at 300 K and 378 K. Further work is under progress to understand the origin of these magnetic transitions.

All magnetic measurements were carried out with specimens oriented 90° in respect of the direction of external magnetic field. It becomes evident, the missing characteristic signature in the shift of the center of magnetic hysteresis loop from its normal position at H = 0 to $H_E \neq 0$, accompanied by an increase in the coercivity (HC) and absence of shift in curie temperatures observed in M-T curve that is between the curie temperatures of the bilayer materials [17], we conclude no exchange interaction was observed between $Fe_3O_4$ - $Gd_5Si_4$ bilayer film.

## Conclusion

In summary, we have successfully deposited $Fe_3O_4$ and $Fe_3O_4$-$Gd_5Si_4$ bilayer thin films of various thicknesses on Glass/Si (1 0 0) substrate by RF Magnetron sputtering technique. Our results indicate magnetic property is dependent on substrate and thickness of the deposited material. There is an observation of high porosity and surface roughness in both the films. No exchange interaction was observed between $Fe_3O_4$-$Gd_5Si_4$ bilayer film. Further improvement in the deposition process is necessary to reduce surface roughness and promote larger physical contact to enhance magnetic interaction between the two layers.